# Simultaneous mapping of the ultrafast time and fluence dependence of the laser-induced insulator-to-metal transition in magnetite

J. O. Schunck[1,2], P. S. Miedema[1], R. Y. Engel[1,2], S. Dziarzhytski[1], G. Brenner[1], N. Ekanayake[1], C.-F. Chang[4], P. Bougiatioti[5], F. Döring[5], B. Rösner[5], C. David[5], C. Schüßler-Langeheine[6] and M. Beye[1,2,3,*]

[1] Deutsches Elektronen-Synchrotron DESY, Notkestr. 85, 22607 Hamburg, Germany

[2] Physics Department, Universität Hamburg, Luruper Chaussee 149, 22761 Hamburg, Germany

[3] Department of Physics, AlbaNova University Center, Stockholm University, SE-10691 Stockholm, Sweden

[4] Max Planck Institute for Chemical Physics of Solids, Nöthnitzer Straße 40, 01187 Dresden, Germany.

[5] Paul Scherrer Institut (PSI), Forschungsstraße 111, 5232 Villigen, Switzerland

[6] Helmholtz-Zentrum Berlin für Materialien und Energie, Albert-Einstein-Straße 15, 12489 Berlin, Germany

[*] martin.beye@fysik.su.se

### Abstract

Pump-probe methods are a ubiquitous tool in the field of ultrafast dynamic measurements. In recent years, X-ray free-electron laser experiments have gained importance due to their ability to probe with high chemical selectivity and at atomic length scales. Measurements are typically repeated many thousands of times to collect sufficient statistics and vary parameters like delay or fluence, necessitating that initial conditions are restored each time. An alternative is presented by experiments which measure the relevant parameters in a single shot. Here, we present a time-to-space mapping imaging scheme that enables us to record a range of delays and laser fluences in any single shot of the X-ray probe. We demonstrate the use of this scheme by mapping the ultrafast dynamics of the optically induced insulator-to-metal Verwey transition in a magnetite thin film, probed by soft X-ray resonant diffraction. By extrapolating our results towards the conditions found at X-ray free-electron lasers with higher photon energy, we demonstrate that the presented data could be recorded in a single shot.

# 1 Introduction

The intriguing properties of complex, functional and quantum materials, like high-temperature superconductivity (*1, 2*), colossal magnetoresistance (*3, 4*), strange metallic behavior (*5, 6*), and insulator-metal phase transitions (*7*) have attracted researchers' interest on their quest to engineer and harvest these phenomena for several decades. Excitation with ultrashort laser pulses is a particularly powerful tool to manipulate equilibrium phases or create transient states which do not occur under equilibrium conditions (*8–11*). Usually, time-resolved studies are performed using two ultrashort pulses: the first pulse (the pump) starts a dynamic process, and the subsequent pulse (the probe) is used to observe the change induced in the sample. Due to typically limited signal levels, this pump-probe process is repeated thousands or millions of times. This allows one to collect sufficient statistics while also varying the temporal separation between the two pulses (the delay) and/or the fluence of the pump pulse.

X-ray free-electron lasers (FELs) offer the opportunity to use X-ray pulses for element-selective and atomic-scale probing with high signal levels and femtosecond temporal resolutions. Many FELs today generate pulses using the process of self-amplified spontaneous emission (SASE) (*12–14*), which results in fluctuating pulse properties, like intensity, pulse duration and spectral content. This characteristic motivated the development of techniques for recording a complete set of information for every single shot. In the post-analysis, the data can then be sorted and binned appropriately. Here, simultaneous detection of a range of parameters offers the opportunity to assign recorded data to its respective state, thereby facilitating a study of such systems in the first place. In repeated pump-probe experiments in general, one needs to take care that each pair of pulses yields the same information. This is compromised e.g. by fluctuating pump and probe beams or jitter between the two. Furthermore, the sample needs to return to the ground state between the pulse pairs and the pulse energies should not cause permanent damage to the sample. Sometimes though, it is interesting to actually probe non-reversible excitations. In this case or if the sample is destroyed, a new sample or spot on the sample needs to be used for every measurement, posing requirements on the available amount or size of sample. Many of these challenges can be alleviated using approaches that can perform a complete experiment, i.e., measure the required delay (and potentially fluence) dependence, in a single shot.

One possible solution is an experimental setup which is able to map a range of one or more parameters simultaneously. In the context of X-ray spectroscopy at synchrotrons, this has for example been done by mapping a range of incident and emitted photon energies onto an area detector (*15–17*). At FELs, one example of the mapping of pump-probe delay range is the X-ray streaking method (*18–20*), used at X-ray free-electron lasers to probe a delay range of approximately 1.5 ps in a single measurement Also different approaches using non-colinear beams (*21–24*) have been realized.

Here, we present the results of setup that enables the simultaneous recording of a delay range of seven picoseconds and a relative fluence range of a factor of more than five. We use an X-ray optical Fresnel zone plate to image the iron $L_3$-edge resonant soft X-ray diffraction (RSXD) signal from a magnetite sample. A non-collinear pump-probe geometry with an angle of 73.25° between the pump and probe beams allows the pump-probe delay to be mapped onto a spatial axis on the detector (see Figure 1). The orthogonal axis images the signal across the laser beam spot onto the detector, essentially mapping the pump laser fluence distribution. The concept of mapping a range of fluences by using zone plates have previously been reported (*25, 26*). With our setup, a large parameter range can be acquired in a static geometry without scanning. Given a sufficiently high signal, such a setup even allows capturing a complete dataset in a single shot.

As a sample, we chose a thin magnetite ($Fe_3O_4$) film. Magnetite continues to serve as a model system for how charge, orbital and lattice degrees of freedom shape an insulator-metal transition, here called Verwey transition (*27–40*): Below 124 K, the resistivity increases by two orders of magnitude due to a structural transition from a high-temperature cubic inverse-spinel structure to a distorted monoclinic structure (*41*), characterized by charge and orbital order (COO) (*35, 36, 38–40*). The excitation of the magnetite low-temperature phase with ultrashort infrared (IR) laser pulses destroys the COO superstructure and drives the system into the high-temperature structure within several picoseconds (*42–47*). Previous time-resolved IR pump, RSXD probe studies (*42, 43*) on magnetite bulk single crystals observed a two-step process on femto- and picosecond time scales, characterized by an initial sub-picosecond reduction of the scattering intensity caused by direct excitation of charge transfer between charge ordered sites (*33, 44*).

## 2 Methods

As outlined in the introduction, a non-colinear geometry of FEL and pump laser is realized in our time-to-space mapping setup, schematically sketched in Figure 1: Due to the angle between the two beams, the relative pulse front arrival time of both pulses varied along the horizontal dimension, thereby mapping a time (i.e. relative delay) axis to the horizontal space coordinate of the sample. With an off-axis Fresnel zone plate (FZP), the resonant diffraction signal from the sample, including this delay axis, was spatially resolved by imaging onto a two-dimensional charge coupled device (CCD). The delay range which can be covered in a single shot was determined by the geometry of the experiment (incident angles of FEL and laser beams), as well as the magnification of the FZP and the properties of the detector. In the present setup, the delay axis was mapped on the CCD as 23.6 fs/pixel. In principle, this would allow us to record a delay range of up to 50 ps, but in our experiment, the simultaneously recorded delay range was limited by the horizontal sample size to approximately 7 ps. Like in a conventional pump-probe scheme, the recorded delay range can be further extended by scanning the pump laser delay stage ($\Delta_{DS}$ in Figure 1).

Both the FEL and the probe laser illuminate an extended area on the order of one square millimeter on the sample, necessitating a homogeneous sample with flat high-quality surface is required. Here, a high-quality magnetite thin film was used. Besides being flat, thin films epitaxially grown on spinel $Co_2TiO_4$ substrates that closely match the crystal structure and lattice constant of magnetite have been shown to have excellent properties (*48*): The resistance hysteresis width of the Verwey transition matches that of the bulk, the correlation length of the COO is comparable to bulk single crystals, and the Verwey transition temperature even surpasses that of the bulk by 3 K.

In order to keep the laser profile rather homogeneous along the probed horizontal direction, which is used as delay axis, we used a cylindrical lens to focus the optical laser to a horizontal line on the sample. In the vertical direction, the laser is focused much stronger, well below the size of the probing FEL beam, such that the full vertical spatial laser profile was probed simultaneously. In the data analysis, we could thus separate regions in the center of the vertical laser spot profile, which were pumped with a high local fluence, from more weakly pumped regions further away.

Our imaging setup recorded two spatial dimensions simultaneously. We used our special experimental configuration to map the spatial coordinates onto one temporal and one local fluence axis. As this scheme introduces a correlation of signals recorded with different delays and fluences with sample position, it works best with samples that are spatially homogeneous over the probed area. If dynamics of a non-homogeneous samples are of interest, our setup has the advantage of parallelizing the data acquisition in comparison to a conventional pump-probe scheme. In such a conventional scheme, inhomogeneities would be studied by focusing the FEL to a spot as small as possible (with damage-related limitations to the usable total number of photons). Subsequently, a full data set would require a four-dimensional scan of pump-probe delay, pump laser fluence and the two lateral dimensions of the sample. In this case, our setup reduces the required scanned dimensions from four to the two spatial dimensions. The ability to study inhomogeneities is closely tied to the spatial resolution of our setup, which we estimate to be around 3 µm to 4 µm for the present geometry (*16*) (see also the experimental section). In the present experiment we scanned the relative pump-probe delay $\Delta_{DS}$ (see Figure 1) back and forth by 3 ps using a movable delay stage, thereby moving the entire 7 ps delay window horizontally over the sample surface. This scan also served as a cross-calibration of the delay axis (see below and Figure 2).

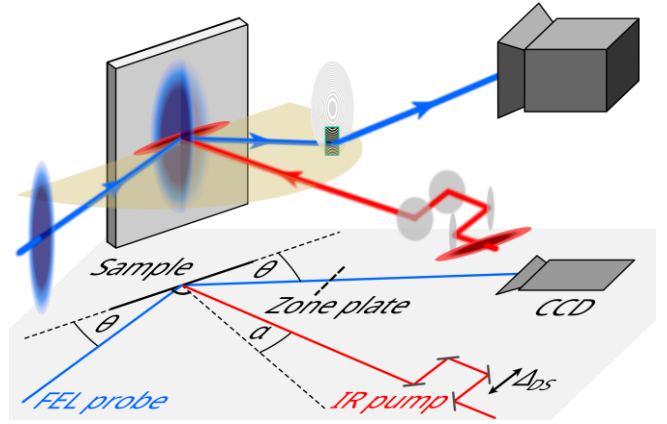

Figure 1: **Time-to-space mapping setup.** Soft X-ray pulses from FLASH resonant to the Fe $L_3$ edge (blue) probe the resonant scattering of $Fe_3O_4$ upon pumping with 800 nm pulses (red). The horizontal scattering plane is shown in yellow. The incidence and scattering angle $\theta$ is 32.75° with respect to the sample surface and the pump laser is incident onto the sample under an angle $\alpha$ of 16° with respect to the sample surface normal. Due to the relative angle between FEL and pump, a delay range of several picoseconds is probed simultaneously. An additional delay $\Delta_{DS}$ can be set by a mechanical delay stage. An off-axis Fresnel zone plate images the resonant scattering of the unfocused FEL beam from the sample onto a two-dimensional CCD detector.

## 3 Experimental setup

Measurements with the time-to-space mapping setup were performed at beamline BL2 (*49–51*) of the free-electron laser (FEL) FLASH at DESY in Hamburg, Germany using the ultrahigh-vacuum diffractometer MUSIX (*52*). The schematic experimental setup is shown in Figure 1. To record resonant diffraction from a magnetite sample at the iron $L_3$-edge at FLASH, third harmonic radiation with a central photon energy of 706 eV was used. The full width at half-maximum (FWHM) bandwidth of the FEL pulses was approximately 3.6 eV. The pulse energy of the FEL fundamental was on average 20 µJ and the third harmonic emission is usually two to three orders of magnitude weaker (*53–55*).

The unfocused FEL beam had a diameter of approximately 4.2 mm at the sample position. The pump laser (*56*) wavelength was centered around 800 nm, the pulse duration was 57 fs FWHM and the pulse energy was 16.8 µJ. The combined temporal resolution of the experiment was ~120 fs (see supplementary material). The laser spot measured approximately 2.6 mm (horizontal, FWHM) by 0.2 mm (vertical, FWHM) and was focused such that a flat-top region formed in the horizontal direction (see supplementary material, Figure S1). The sample was horizontally smaller than the laser footprint, extending only in the flat-top region of the laser focus. The pulse repetition rate of the FEL was 10 Hz, while that of the pump laser was 5 Hz so every other FEL shot probed the unpumped sample, serving as a reference for normalization of the signal. We verified that the sample was fully recovered in between shots. Due to the small third harmonic fraction of the FEL beam, the scattering cross-section of the sample and the large FEL footprint, the count rate for the data set was on average 0.54 counts/shot. In total, the data shown here comprises 12 h of acquisition time and approximately 230,000 counts.

The pump laser beam was incident under an angle of 16° with respect to the sample surface normal (see Figure 1). The sample was a 40 nm thin magnetite ($Fe_3O_4$) film, grown on $Cr_2TiO_4$ (001) (*48*), measuring approximately 1.9 mm (horizontal) by 8 mm (vertical). The incidence angle $\theta$ of the FEL was 32.75° with respect to the sample surface to fulfill the Bragg condition for the (00½) superstructure peak. The sample was cooled to a base temperature of 100 K for the entire duration of the measurements. An off-axis Fresnel zone plate (FZP) imaged the diffraction signal onto a two-dimensional charge-coupled device (CCD), thereby magnifying the diffraction signal from the sample by a factor of 4.1. Further details on the experimental setup can be found in the supplementary material.

## 4 Results

As described in the methods section, we obtained data as a function of the horizontal sample position, $x$, the vertical sample position, $y$, and the delay stage position, $\Delta_{DS}$. How we obtain the calibration of the delay axis on the detector coordinates from this multidimensional data set, as well as data treatment steps used to obtain fluence-dependent delay traces, are visualized in Figure 2. Small regions of the sample imaged at delay stage positions $\Delta_{DS}$ of $0 \pm 0.5$ ps and $2 \pm 0.5$ ps are shown in panels (a) and (b), respectively. In both panels, the depicted pump signal has been normalized to the unpumped signal. The part of the sample where the COO was pumped away by the laser is visible as blue area. The horizontal position where both pulses cross the surface at the same time (time overlap, $t_0$) can be seen to be the $x$ position where the signal transitions from high to low scattering intensity (black dashed lines). We chose the origin of the $\Delta_{DS}$ scale such that $t_0$ is in the horizontal center of the detector window ($x = 150$ px), see Figure 2(a).

The vertical profile of the signal which is averaged in the pumped area of panel (b) along the $x$-axis is shown as turquoise data in panel (c) of Figure 2. This profile maps the local strength of the pump-probe effect and follows the local fluence profile of the laser, which we show overlaid to the pump laser profile as extracted from an independent knife edge scan (green).

We now focus on calibrating the delay scale. For this, we use the region with the largest pump effect in the vertical center of the laser spot profile. We average the $x$-traces with pump fluences higher than 4 mJ/cm² and plot these for the different delay stage positions as a map (Figure 2(d)). $t_0$ extends along a diagonal (dashed line) reflecting the linear influences of both $x$ and $\Delta_{DS}$ on the delay. The slope of this diagonal line yields the delay mapping factor on our detector (here 23.6 fs/px). This value depends on the relative incidence angle between FEL and pump laser, the sample and observation angles, the magnification of the imaging zone plate and the pixel size of the detector. By choosing these parameters, the setup-limited temporal

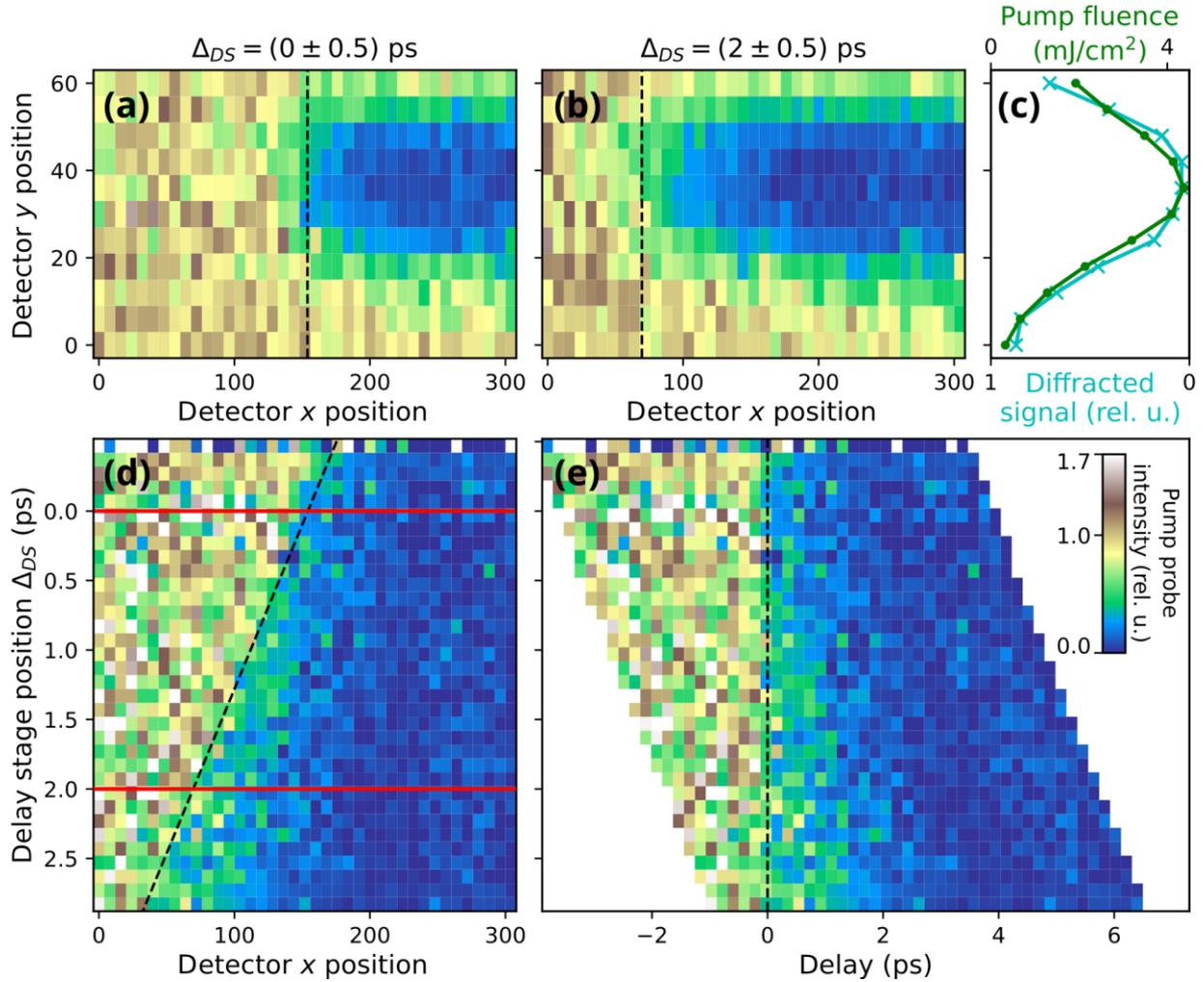

Figure 2: **Pump-probe data treatment.** The pump-probe delay varies along the horizontal *x*-axis because of the different incidence angles of FEL and pump laser. Panels (a), (b): Normalized images of the pump effect on the sample obtained by dividing pumped by unpumped events for delay stage position ranges of $\Delta_{DS} = (0 \pm 0.5)$ ps and $\Delta_{DS} = (2 \pm 0.5)$ ps, respectively. One positional unit on the detector corresponds to 13.5 µm. Panel (c) shows the laser beam profile (green, upper axis) and the corresponding profile of the diffracted signal (cyan, lower axis), obtained by averaging only the pumped signal in panel (b) along the horizontal axis. Panel (d): By scanning the pump laser delay stage position $\Delta_{DS}$ (see also Figure 1) and plotting the average of detector *y* positions pumped with the highest fluences against the delay stage position (data shown along the red lines are extracted from panels (a) and (b)) allows one to observe how temporal overlap moves along *x* (black dashed line). The slope of the black dashed line can be used to translate every row onto a common delay axis. The *x*-axis and *y*-axis have been binned by a factor of 8 and 6, respectively. Panel (e) then shows the data set on the common horizontal delay axis, while the vertical axis still shows the mechanical delay stage position. Pump-probe traces as shown in Figure 3 are finally derived by averaging the data in panel (e) along the vertical dimension.

resolution can be tuned to match the experimental requirements.

With this conversion, we can align each row to a common delay axis, see Figure 2(e). A common delay trace using all available data is then obtained by averaging the data in Figure 2(e) along the *y* axis. We note that there is a hint towards a small sample position dependence of the observed pump effect; it appears that the sample region that is mapped onto $x \approx 90$ exhibits a somewhat slower pump dynamics, which can be seen as a comparatively higher intensity around a delay value of 1 ps for delay stage positions of approximately 2.5 ps. However, the present data set does not have sufficient quality to study this rather weak effect in more detail.

Usually, the pump-fluence dependence of the ultrafast response of the sample to IR excitation is determined by repeating the pump-probe experiment with different pump laser attenuation settings. In our setup, we can do this by sorting different rows of the detector image according to the pump fluence at that position, see Figure 2(c). We chose to sort the data into five fluence bins (see supplementary material; Figure S2). Thus, we obtain five delay traces spanning a fluence range between 0.7 mJ/cm² and 4.1 mJ/cm² without actually changing the incident laser pulse energy. The resulting fluence-dependent IR pump, RSXD probe traces are depicted in Figure 3(a), showing a double exponential decay characterized by a fast and a slow component.

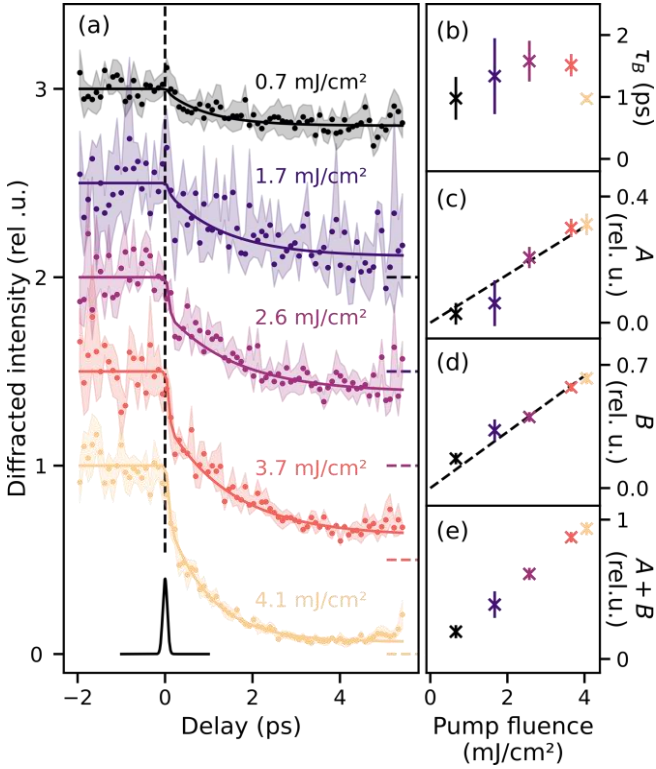

Figure 3: **Ultrafast fluence-dependent response of a 40 nm thin magnetite film to 800 nm laser excitation.** Panel (a): The intensity of the (00½) superstructure peak decreases on a few-picosecond time scale upon irradiation with the pump laser for fluences between 0.7 mJ/cm² and 4.1 mJ/cm². The data was normalized to 1 for negative delays and fit curves were offset by 0.5 relative units (rel. u.) for clarity, and the dashed lines to the right show 0 intensity for the data of the respective fluence. The shaded area is the one-sigma measurement uncertainty of the data, calculated as the standard deviation for each delay value upon averaging several traces of similar fluence and binning on the delay axis. Full lines show fits with a double exponential decay convolved with a Gaussian profile with a width corresponding to the experimental resolution, shown below in black at a delay of 0 ps. The fast time constant $\tau_A$ is set to be much faster than the temporal resolution. The fluence dependence of the slower time constant $\tau_B$ is shown in panel (b). Panels (c) and (d): Fluence dependence of the amplitudes $A$ (fast process) and $B$ (slow process) alongside linear trend lines with slopes 0.08 rel. u./(mJ/cm²) and 0.16 rel. u./(mJ/cm²), respectively. Panel (e): The total decay amplitude (i.e. sum of $A$ and $B$). See supplementary material for details on the fitting procedure.

The resulting traces were fitted using a double exponential decay function (solid lines) convolved with a Gaussian of a width corresponding to the temporal resolution of the experiment (120 fs FWHM) as a fit model. The shaded regions in Fig. 3(a) show the experimental one-sigma standard deviation calculated for each data point during the averaging of multiple traces and binning on the *x*-axis. During the weighted least-squares fitting, the inverse experimental error is used as weights for the respective data points. We find that following the IR excitation, a fast and a slow decay process, respectively named $A$ and $B$, occur, matching what has previously been observed for single crystals (*42*, *43*, *57*). Within our experimental uncertainty, there are no indications of a recovery in the measured delay range.

In panels (b) through (e) of Figure 3, the fluence dependence of the fit results for the slow decay time constant $\tau_B$ (b), the amplitudes $A$ (c) and $B$ (d) and the total long-time decrease $A+B$ (e) are shown. Error bars represent the one-sigma standard deviations. The fast decay is limited by the temporal resolution of the experiment to 120 fs. In the fitting procedure, this is realized by the convolution with the resolution-limited Gaussian function, therefore the value of $\tau_A$ was fixed to a value of 10 fs, much faster than the temporal resolution. Both amplitudes, $A$ and $B$, show a linear increase with fluence (black dashed lines in Figure 3(b) and (c)). The slow decay time constant $\tau_B$ ranges from 1 ps to 2 ps, generally matching previously reported values (*43*). See the supplementary material for more details on the fitting procedure.

## 5 Discussion

In the data shown before, we analyze the temporal evolution of the intensity of the (00½) superstructure peak, which is a direct measure for the degree of charge and orbital order in magnetite. In the low-temperature ordered phase, this reflection appears in resonant soft X-ray diffraction (RSXD) e.g. at the iron $L_{2,3}$-edges and disappears when the sample transforms into the high-temperature phase, indicating the disappearing charge and orbital order (*29*, *30*, *39*, *40*, *58*). The IR pump laser can drive this phase transition and thus the (00½) reflection vanishes as a function of pump-probe delay. The IR pump, RSXD probe results obtained on magnetite thin films qualitatively match those previously reported for single crystals (*42*, *43*, *57*): we find a biexponential decay with a fast time constant $\tau_A$ limited by the 120 fs temporal resolution and a slow time constant on the picosecond scale. The linear fluence dependence of amplitude *A*, previously reported for single crystals (*43*), is confirmed by our measurements. Quantitative differences beyond these generally agreeing results are attributed to differing experimental conditions: The fact that we do not observe two distinct fluence regimes is because the lowest fluence shown in Figure 3a is already sufficient to drive the sample from its initial temperature of 100 K through the Verwey transition at 124 K. With a higher signal level, our results could have been binned finer to allow drawing conclusions also on the lower fluence regime. Furthermore, the fluence dependence of the fast decay amplitude *A* has been reported to scale as 0.3 rel.u./(mJ/cm$^2$) for single crystals (*43*), while we find 0.08 rel.u./(mJ/cm$^2$) for the thin-films studied here.

This dependency likely arises because a comparison of fluences in different experiments is challenging. Small deviations coming from differing calibrations in spot sizes and laser pulse energies, different estimates of sample reflectivity (that also relate to surface quality) as well as using different incidence angles and polarizations may add up to sizeable differences. For a quantitative comparison, one would ideally study both sample types in the same setup. As intrinsic sample properties are concerned, differences between our magnetite thin film and a bulk magnetite crystal are expected to be small: A study directly comparing static RSXD and X-ray absorption data of thin films and bulk single crystals only finds minor spectral differences (*58*). The limited film thickness is likely to not explain this behavior, since the X-ray probing depth of around 10 nm to 40 nm (*42*, *43*) is shorter than the film thickness. Different thermal conductivities of the film substrate as compared to the bulk sample are expected to influence the evolution only on longer timescales than studied here.

Having discussed the observed dynamics in magnetite, we now turn to our time-to-space mapping method. In its current implementation, signal levels of the setup were limited to approximately 0.5 counts/shot on average. Within an acquisition time of 12 h, we collected about 230,000 shots from the FEL, comprising the data shown in Figure 3. Mainly two aspects contribute to this rather low count rate: The used pulses from FLASH had 20 μJ pulse energy in the fundamental with an expected contribution at the Fe *L*-edge of about 20 nJ. Additionally, focusing the FEL beam to a round spot with about 4.2 mm diameter results in a loss of photons of more than one order of magnitude, if compared to the actually relevant photons probing the pumped area of the sample which measured $2 \times 0.2$ mm$^2$. Improving these two aspects could thus gain a more than four orders of magnitude in signal. If additionally higher-energy FELs, like the European XFEL, PALFEL, SwissFEL or LCLS, which can typically produce mJ level pulse energies at these photon energies (*59*), were used, a total gain of about six orders of magnitude could be realized in an optimized experiment. With these improvements, the signal level of a single shot has the potential to substantially surpass that of the full data set presented here meaning that, our setup can be used in the future to record full datasets on laser-induced dynamics of sensitive samples in a single shot with complete information on time- and fluence dependences.

We note though that the conditions described above result in probe fluences of about 500 mJ/cm$^2$, which is certainly beyond the damage threshold of the majority of solids. Consequently, single-shot acquisition becomes a necessity in a measure-before-destroy approach. In order to reduce the probe fluence to a level of <1 mJ/cm$^2$, which should be tolerable and non-disturbing to most samples (*60*), the FEL pulse could be attenuated by about one order of magnitude while still yielding a data set of comparable quality to that shown here. The remaining about two orders of magnitude can be recovered by increasing the probe spot size, for example, by using an FEL spot size of $6 \times 6$ mm$^2$. While solid samples and sample holders of this size are not uncommon, this size requirement may pose a challenge if e.g. homogeneous magnetic fields are required to prepare certain sample states. An X-ray focus of this size is routinely achievable with bendable KB-optics which are relatively common at modern FEL facilities. Consequently, the spot size of the optical pump laser would need to be increased accordingly, necessitating an increase in pulse energy to adequately excite the sample.

If these requirements can be fulfilled, the current delay and fluence ranges which are probed simultaneously can be increased. In its current implementation, the setup probes a delay range of 7 ps and a fluence range of more than a factor of five in a single acquisition. The delay range depends on the size of the horizontal pump laser-FEL overlap and the angle between the two. An overlap range of 6 mm, as discussed above, would increase the observed delay range to 21 ps. The current distribution of fluence bins (see Figure 3 and the supplementary material) is currently restricted by the limited signal level. Improvements in signal would enable a finer binning of the laser fluence profile and

## 6 Conclusion and Outlook

In this study, we present results of a time-to-space mapping setup for time-resolved pump-probe experiments. With this setup, the ultrafast phase transition dynamics of high-quality magnetite ($Fe_3O_4$) thin films was investigated, induced by 800 nm laser pulses using resonant soft X-ray diffraction at the iron $L_3$-edge as probe. The induced dynamics generally match those previously observed in single crystals.

This setup is capable of recording a range of delays (dependent on the dimensions of the sample and the laser and FEL beams) and fluences in a static setup, without the need for any scans. This is achieved within a range of a few millimeters on the sample where the pump laser and FEL overlap, both incident under a relative angle of 73.25°. Due to this angle, the relative delay of both pulses is imprinted into the horizontal dimension on the sample and imaged onto the detector with a Fresnel zone plate. In the vertical direction, imaging the laser spot profile allows for the simultaneous probing of regions of the sample that have been pumped by higher (in the center of the spot) or lower (at the edges of the spot) fluences.

Besides its use for resonant scattering experiments, the method presented in this work for mapping of delay and fluence ranges can be transferred to other X-ray spectroscopy schemes, like non-resonant scattering, X-ray absorption measured in transmission or reflectivity, and potentially X-ray magnetic circular dichroism. The resonant diffraction efficiency of magnetite at the iron $L_3$-edge is about $10^{-4}$ (*57*), which is small in comparison to e.g. absorption measured in transmission. Since the necessary sample and spot sizes as well as pulse energies critically depend on the signal level, other detection modalities will change the feasibility of this approach (see discussion section).

Acquisition schemes similar to our method could also be realized in laboratory-based experiments using high-harmonic generation (HHG). While such a source would not provide enough photons for single-shot acquisitions, the stability together with our non-scanning approach may yield excellent data quality after averaging.

also allow to increase the probed fluence range to lower fluences.

## 7 Supplementary Material

The supplementary material provides some more detailed information on the calibration of the pump laser fluences, experimental parameters (temporal resolution of the setup, Fresnel zone plate optic and detector) and the fitting procedure of the pump-probe delay traces.

## 8 Acknowledgments

The setup as well as MB, PSM, RYE and JOS are funded through a grant to MB for a Helmholtz young investigator group under contract number VH-NG-1105. CSL acknowledges funding by the Deutsche Forschungsgemeinschaft (DFG, German Research Foundation)– Project-ID 328545488 – TRR 227. We acknowledge DESY (Hamburg, Germany), a member of the Helmholtz Association HGF, for the provision of experimental facilities. Parts of this research were carried out at FLASH and we would like to thank Sven Toleikis for helpful discussions and assistance in using beamline BL2, as well as the DESY machine and photon operators for assistance in using the FEL beam. Beamtime was allocated for proposals F-20170534 and F-20181193.## 9 Author Declarations

**Conflicts of interest**

The authors have no conflicts to disclose.

**Author contributions**

**J.O.S.** Formal Analysis (lead), Investigation (equal), Visualization (lead), Writing/Original Draft Preparation (equal), Writing/Review & Editing (lead). **P.S.M.** Investigation (equal), Writing/Review & Editing (equal). **R.Y.E.** Formal Analysis (supporting), Investigation (equal), Writing/Review & Editing (equal). **S.D.** Investigation (equal), Writing/Review & Editing (equal). **G.B.** Investigation (equal), Writing/Review & Editing (equal). **N.E.** Investigation (equal), Resources (equal), Writing/Review & Editing (equal). **C.-F.C.** Resources (equal), Writing/Review & Editing (equal). **P.B.** Investigation (equal), Writing/Review & Editing (equal). **F.D.** Resources (equal), Writing/Review & Editing (equal). **B.R.** Investigation (equal), Resources (equal), Writing/Review & Editing (equal). **C.D.** Resources (equal), Writing/Review & Editing (equal). **C.S-L.** Conceptualization (equal), Investigation (equal), Writing/Original Draft Preparation (equal),

Writing/Review & Editing (equal). **M.B.** Conceptualization (lead), Funding Acquisition (lead), Investigation (equal), Project Administration (lead), Supervision (lead), Writing/Original Draft Preparation (equal), Writing/Review & Editing (equal).

## 10 Data Availability Statement

The data that supports the findings of this study are available from the corresponding author upon reasonable request.